\documentclass[aps,nofootinbib]{revtex4}
\usepackage{epsfig}
\usepackage[latin1]{inputenc}
\usepackage{float,amsmath,amstext,amsthm,amssymb,amsfonts,slashed}
\usepackage{graphicx}

\begin{document}

\title{Dilepton production from non-equilibrium hot hadronic matter}

\author{B.~Schenke$^{1}$ and C.~Greiner$^{1}$}

\affiliation{$^{1}$Institut f\"ur Theoretische Physik, %
   Johann Wolfgang Goethe Universit\"at Frankfurt, %
   Max-von-Laue-Str.~1, %
   D--60438 Frankfurt, %
   Germany \vspace*{5mm}}

\begin{abstract}

It is investigated under which conditions an adiabatic adaption of the dynamic and
spectral information of vector mesons to the changing medium in heavy ion collisions,
as assumed in schematic model calculations and microscopic
transport simulations, is a valid assumption.
Therefore time dependent medium modifications of low mass vector mesons
are studied within a nonequilibrium quantum field theoretical description.
Timescales for the adaption of the spectral properties are given and
non-equilibrium dilepton yields are calculated, leading to the result
that memory effects are not negligible for most scenarios.

\end{abstract}

\maketitle


\section{Introduction and Motivation }
\label{Intro}

High energy heavy ion reactions allow for studying strongly interacting matter under extreme
conditions, i.e., high densities and temperatures. Photons and
dileptons do not undergo strong interactions and thus may carry
undistorted information on the early hot and dense phases of the
fireball, because the production rates
increase rapidly with temperature. Dilepton
spectra are expected to play a central role in inferring
the restoration of the spontaneous breaking of
chiral symmetry from heavy ion reactions. In the low
mass region they couple directly to the light vector mesons and
reflect their mass distribution. They are thus considered the
prime observable in studying mass (de-)generation related to the
restoration of spontaneous chiral symmetry breaking.

In this work we investigate dilepton production from the hot
hadronic medium and we will
concentrate on the low mass region, particularly the medium
modifications of the $\rho$-
meson. The medium and hence also possibly the properties of the
regarded mesons undergo substantial changes over time. Such
scenarios have been described within transport calculations using
some quantum mechanically inspired off-shell propagation
\cite{Effenberger:1999ay,Cassing:1999wx}. It emerges the important
question, whether a quasi instantaneous adaption of the dynamic
and spectral information to the changing medium, as assumed in
more schematic fireball model calculations \cite{Rapp:1999ej} and
microscopic transport simulations, is a suitable assumption or
whether the vector meson's spectral information reacts to changes
with a certain "quantum mechanical" retardation. The transport
description of off-shell excitations is an open field of research
and necessary in order to understand the transport dynamics of
resonances. We employ a nonequilibrium quantum field theoretical
description based on the formalism established by Schwinger and
Keldysh \cite{Schw61,Ke64}. We give a formula for the dynamic
dilepton production rate and simulate modifications of the light
vector mesons due to the dynamically changing medium in heavy ion
collisions by parameterizing a certain time dependence of the
$\rho$-meson self energies. We are able to analyze the mesons'
dynamic spectral properties as well as the resulting dilepton rate
and the yield from an evolving fireball and compare to the
quantities computed assuming adiabaticity.

\section{The nonequilibrium production rate}
\label{dileptonproduction}
    We utilize the Schwinger-Keldysh formalism
    in order to derive the dynamic non-equilibrium rate of produced
    electron-positron pairs, coming from the decay of light vector
    mesons via virtual photons in a spatially homogeneous system (details in \cite{sg05}).

    Projecting on the particle number in the electron propagator $G^{<}$
    and using the equations of motion for $G^<$, the so called Kadanoff-Baym equations,
    we find the production rate of electrons for a homogeneous, yet time dependent system, to read

    \begin{align}
        \partial_{\tau} N(\textbf{p},\tau)
        &=2 ~ \text{Im}\left[ \text{Tr}\left\{\frac{\slashed{p}+m}{2 E_{\textbf{p}}}\int_{t_0}^{\tau}d\bar{t} \left(\Sigma^<(\textbf{p},\tau,\bar{t})\right)
        e^{i E_{\textbf{p}}(\tau-\bar{t})}\right\}\right]\text{,}\label{pertfinal}
    \end{align}
    with the electron self energy $\Sigma$ and $p_0 = E_{\textbf{p}}$.
    The free electron propagator can be used because due the electrons' long mean free path the electrons are not expected to interact with the medium after being produced.

\begin{figure}[H]
      \begin{center}
        \includegraphics[width=3cm]{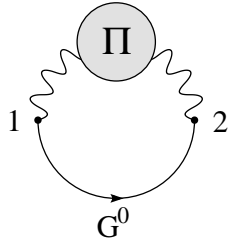}
        \caption{$\Sigma(1,2)$ in coordinate space}
        \label{fig:selfenergy}
      \end{center}
    \end{figure}

    The medium effects enter via the dressing of the virtual photon propagator
    in the electron self energy (see Fig. \ref{fig:selfenergy}).
    ~\\
     $D_{\gamma}$ is the virtual photon propagator and $\Pi$ is its self energy. We have
    \begin{align}
        i \Sigma^{<}(\textbf{p},t_1,t_2)= - e^2
        \gamma_{\mu}\left(\int
        \frac{d^3k}{(2\pi)^3}D_{\gamma}^{<,\mu\nu}(\textbf{k},t_1,t_2)G^{<}_{0}(\textbf{p}-\textbf{k},t_1,t_2)\right)\gamma_\nu\text{,}
    \end{align}
    with $\textbf{k}$ the momentum of the virtual photon.
    On inserting this self energy and
    defining $p^{+}=k-p$ as the four-momentum of the positron and $p^{-}=p$ as that of the electron,
    equation (\ref{pertfinal}) becomes
    \begin{align}
        E_+E_-\frac{dR}{d^3p^{+}d^3p^{-}}(\tau)=&\frac{2e^2}{(2\pi)^6}
        \left[p_{\mu}^+p_{\nu}^-+p_{\nu}^+p_{\mu}^--g_{\mu\nu}(p^+p^{-}+m^2)\right]
        \text{Re}\left[\int_{t_0}^{\tau}d\bar{t} i
        D_{\gamma}^{<,\mu\nu}(\textbf{k},\tau,\bar{t})e^{i(E_+ +
        E_-)(\tau-\bar{t})}\right]\text{,}\label{galekapustarate}
    \end{align}
    with $E_+=E_{\textbf{p}}$ and $E_-=E_{\textbf{k}-\textbf{p}}$.

    Applying the equilibrium properties of $D^<$, it can be shown that equation (\ref{galekapustarate}) is the generalization of the well
    known thermal production rate for lepton pairs in the stationary case \cite{gk91}:
    \begin{align}
        E_+E_-\frac{dR}{d^3p^{+}d^3p^{-}}(\tau)&=-\frac{2e^2}{(2\pi)^6}
        \left[p_{\mu}^+p_{\nu}^-+p_{\nu}^+p_{\mu}^--g_{\mu\nu}(p^+p^{-}+m^2)\right]\frac{1}{M^4} \frac{1}{e^{\beta E}-1}\text{Im}\Pi_{\gamma}^{\text{ret},\mu\nu}(k,\tau)\label{galekapustarate2}
    \end{align}
    In the following, we will consider the mode $\textbf{k}=0$ exclusively, i.e., the virtual photon resting with respect
    to the medium. After projecting on the virtual photon momentum and taking the electron mass to zero, we get
    \begin{align}
        \frac{dN}{d^4xd^4k}(\tau,\textbf{k}=0,E)=&\frac{2e^2}{(2\pi)^6}\frac{2}{3}\pi(k_{\mu}k_{\nu}-k^2g_{\mu\nu})
        \text{Re}\left[\int_{t_0}^{\tau}d\bar{t} i D_{\gamma}^{<,\mu\nu}(\textbf{k}=0,\tau,\bar{t})e^{i E (\tau-\bar{t})}\right]\text{.}
        \label{sigmalessrate}
    \end{align}
    The dynamic information is inherent in the memory
    integral on the right that runs over all virtual photon occupation numbers $D_{\gamma}^{<}$ from the initial time to
    the present. This way the full nonequilibrium electron production rate at the present time $\tau$ is determined.
    We introduce the dynamic medium dependence by dressing the virtual photon propagator
    with the medium dependent $\rho$- or $\omega$-meson. This
    dressing enters with the self energy $\Pi^<$ via the relation
    \begin{align}
        D^{\gtrless}(1,1')=&\int_{t_0}^{\infty}d2\int_{t_0}^{\infty}d3D^{+}(1,2)\Pi^{\gtrless}(2,3)D^{-}(3,1')+\text{surface term,}
        \label{fdt}
    \end{align}
    that follows from the Kadanoff-Baym equations.
    Using projectors on the different polarizations and the equality of each polarization ($\textbf{k}=0$), we find
        \begin{align}
           \frac{dN}{d^4xd^4k}(\tau,E,\textbf{k}=0)=&\frac{2}{3}\frac{e^2}{(2\pi)^5}(3E^2)
           \text{Re}\left[\int_{t_0}^{\tau}d\bar{t} i D_{\gamma,T}^{<}(\textbf{k}=0,\tau,\bar{t})e^{i E
           (\tau-\bar{t})}\right]\text{.}\label{photrate1}
        \end{align}
    For dilepton production $\Pi^{\text{ret}}\propto e^2$ and
    $E$ is the invariant mass of the virtual photon. For the
    cases we are interested in, $|\Pi^{\text{ret}}|\ll E$ and we can approximate
$
            D_T^{<}=D_0^{\text{ret}}\otimes\Pi_T^{<}\otimes D_0^{\text{av}}\label{conv0}\text{.}
$
    Diverging contributions at early times (low frequencies), due to the undamped photon propagators, cause numerical problems.
    We introduce an additional cutoff $\Lambda$ for these propagators:
$
        D_0^{\text{ret}}(\tau-t_1) = (\tau-t_1) \rightarrow (\tau-t_1) e^{-\Lambda (\tau-t_1)}
$
    and analogously for $D_0^{\text{av}}(t_2-\bar{t})$.
    The exponential factors lead to a reduction of the rate, which
    we will overcome by renormalizing the final result.
    This does not affect
    the timescales we are interested in, and comparison of the
    dynamically computed rate for a stationary situation (constant
    self energy) with the analytic, thermal rate shows perfect
    agreement.

    Vector meson dominance (VMD) allows for the
    calculation of the photon polarization tensor $\Pi_T^<$, using the identity between the electromagnetic
    current and the canonical interpolating fields of the vector mesons,
    which leads to
$
            \Pi^<_{\alpha\beta}=\frac{e^2}{g_\rho^2}m_{\rho}^4
            D_{\rho, \alpha\beta}^{<}\text{.}
$
    Once more, we apply the generalized fluctuation dissipation
    relation (\ref{fdt}) to calculate
        \begin{align}
            D_{\rho,T}^{<}=D_{\rho,T}^{\text{ret}}\otimes\Sigma^<_{\rho,T}\otimes
            D_{\rho,T}^{\text{av}}\text{,}\label{photfdr}
        \end{align}
    with the $\rho$-meson self energy $\Sigma^<_{\rho}$.
    The transverse parts of the retarded and advanced propagators $D_{\rho,T}^{\text{ret}}(\textbf{k},t_1,t_3)=D_{\rho,T}^{\text{av}}(\textbf{k},t_3,t_1)$
    of the vector meson in a spatially homogeneous and isotropic medium follow the equation of motion
        \begin{align}
            \left(-\partial_{t_1}^2-m_{\rho}^2-\textbf{k}^2\right)D_{\rho,T}^{\text{ret}}(\textbf{k},t_1,t_3)-\int_{t_3}^{t_1}d\bar{t}
            \Sigma^{\text{ret}}_{\rho,T}(\textbf{k},t_1,\bar{t})D_{\rho,T}^{\text{ret}}(\textbf{k},\bar{t},t_3)=\delta(t_1-t_3)\text{.}\label{photdgl}
        \end{align}
    The dynamic medium evolution is now introduced by hand via a specified time dependent retarded meson self
    energy $\Sigma^{\text{ret}(\tau, \omega)}$ \cite{sg05}. From that the self energy $\Sigma^<$, needed for solving equation (\ref{photfdr}),
    follows by introducing a background temperature of the fireball.
    The fireball is assumed to generate the time dependent self energy $\Sigma^{\text{ret}}$ and, assuming a nearly quasi thermalized system,
    the $\rho$-meson current-current correlator $\Sigma^<$ is given via $\Sigma^<(\tau,\omega)=2i n_B(T(\tau)) Im \Sigma^{\text{ret}}(\tau,\omega)$
    , which follows from the KMS relation \cite{Greiner:1998vd}, being valid for thermal systems. The latter is a rather strong assumption,
    but necessary in order to proceed.


\section{The medium in nonequilibrium}
The medium effects are introduced via a specific evolving self energy
of the vector meson. A simple self energy
        \begin{align}
            \Sigma^{\text{ret}}(\omega,\tau)=-i\omega\Gamma(\tau)\text{,}
        \end{align}
with a $\textbf{k}$- and $\omega$-independent width $\Gamma$,
leads to a Breit-Wigner distribution for the spectral function.
The time dependence is being accounted for by introduction of the
parameter $\tau$.
For the $\textbf{k}=0$ mode, the full
self energy for coupling to $J^P=\frac{1}{2}^-$ -resonances is
given by
        \begin{align}
            \Sigma_T(\omega,\textbf{k}=0)=\frac{\rho}{2}
            \left(\frac{f_{RN\rho}}{m_{\rho}}\right)^2 g_I
            \frac{\omega^2 \bar{E}}{(\omega+\frac{i}{2}\Gamma_R)^2-\bar{E}^2}-i\omega\Gamma
            \label{modelshenself}\text{,}
        \end{align}
with $\bar{E}=\sqrt{m_R^2+\textbf{k}^2}-m_N$ and $m_R$ and $m_N$
the masses of the resonance and the nucleon respectively.
$\Gamma_R$ is the width of the resonance and $g_I$ the isospin
factor \cite{po04}. For our purpose it suffices to retain the part
of the self energy that creates the pole structure. We will
neglect the $\omega^2$ in the numerator because it causes
straightforward dispersion relations to become invalid (subtracted
dispersion relations are needed in this case). We will shorten
$\left(\frac{f_{RN\rho}}{m_{\rho}}\right)^2 g_I/2~\omega^2$, with
$\omega^2=1 \text{GeV}^2$ by $C$, a dimensionless factor.

    \begin{figure}[H]
      \begin{center}
        \includegraphics[width=7cm]{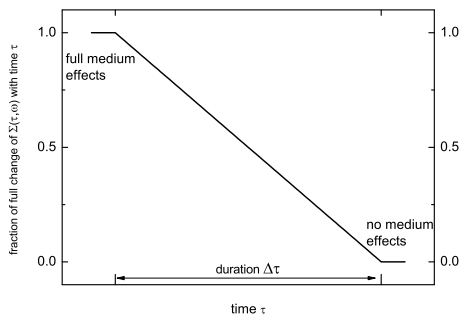}
        \caption{\small Linear switching off of in-medium effects over a certain time $\Delta \tau$}
        \label{fig:linearchange}
      \end{center}
    \end{figure}

Investigating which contributions in equation
(\ref{sigmalessrate}) come from which times in the past, shows
that there are contributions from early times as well as
alternating positive and negative contributions. An interpretation
of this becomes difficult and it follows that only the time
integrated yield is a physical quantity.

\begin{figure}[H]
      \begin{center}
        \includegraphics[height=9cm]{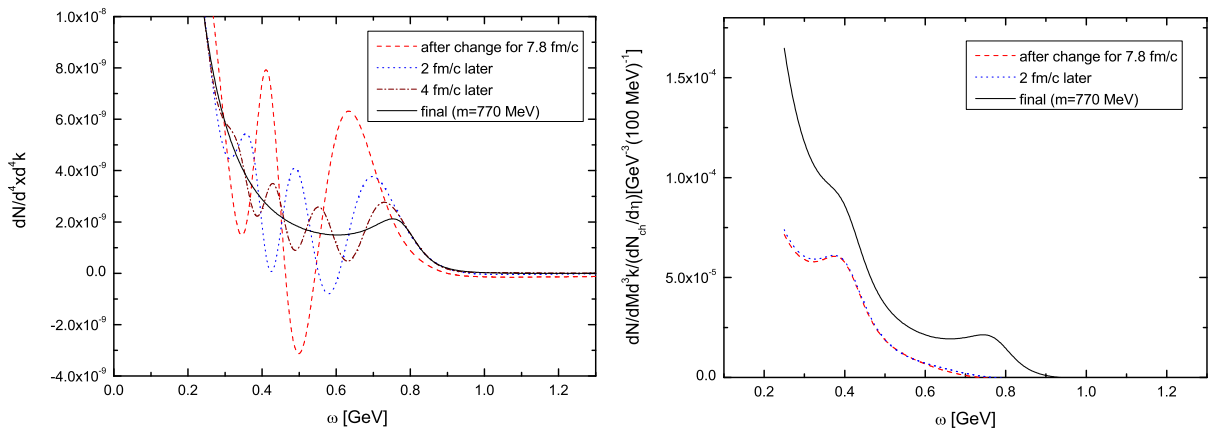}
        \caption{\small Production rate for the change of the mass from $m=400$ MeV to 770 MeV (constant $\Gamma$=150 MeV and constant $T=160$ MeV) directly after the self energy has reached its
                 final form (after 7.8 fm/c) and 2 (and 4) fm/c later. Oscillations and negative values appear in the intermediate rates (left). The corresponding yield stays positive (right).}
        \label{fig:oscillations}
      \end{center}
    \end{figure}

\begin{figure}[hbt]
      \begin{center}
        \includegraphics[height=9cm]{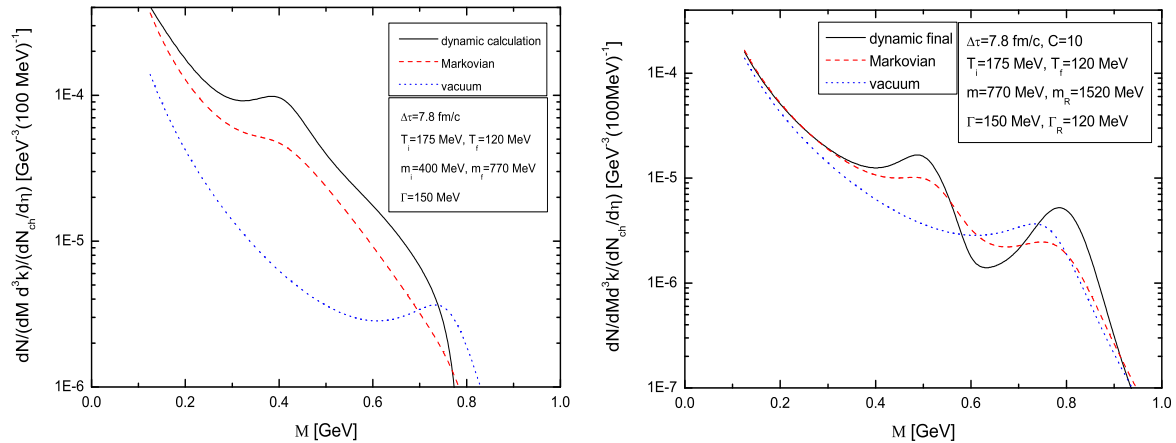}
        \caption{\small Comparison of the dynamically computed dilepton yield from the fireball (solid line) to the one calculated assuming adiabaticty
                 (as done in principle in \cite{Rapp:1999ej})(dashed line),
                 to show the differences caused by memory effects.
                 Mass shift to 400 MeV in-medium mass (left) and coupling to the N(1520) resonance (right).
                 No broadening of either the resonance or the $\rho$-meson is included here. Also shown is the yield coming from a constant
                 $\rho$ meson's vacuum spectral function (dotted line) for comparison.}
        \label{fig:yieldmassres}
      \end{center}
    \end{figure}

In order to quantify the times that the mesons' spectral
properties need to adjust to the evolving medium we change the
self energy linearly in time (see Fig. \ref{fig:linearchange}). As
a possible characteristic timescale we consider the difference of
the final spectral function to the dynamically calculated one at
the time where the medium effects are fully turned off, described
by the difference in the moment $\int_0^{\infty} A(\omega)^2
\omega^2 d\omega$ of the two spectral functions or the difference
in the peak position and height. All methods lead to similar
results \cite{sg05}. We find an exponentially decreasing
difference with increasing duration of the change $\Delta\tau$
(see Fig. \ref{fig:linearchange}), from that we extract a time
constant $\tilde{\tau}$. For the the $\rho$-meson we find a
typical timescale of about 3 fm/c. That means that the spectral
properties follow the changes in the medium nearly instantaneously
only if the evolution is very slow as compared to 3 fm/c. This
means that the vector mesons possess a certain memory of the past,
and even if they decay outside the medium, they still carry
information on the medium in that they were produced. This becomes
important especially for the $\omega$-meson, having a width of
only $8.49$ MeV ($\tilde{\tau}\approx 60$ fm/c). It turns out that
$\tilde{\tau}$ is proportional to $c/\Gamma_2$ with $c$ lying
between $2$ and $3.5$, depending on $m$ and $\Gamma_1$. This is
significantly longer than the naively expected timescale of
$1/\Gamma$. The time needed by the dilepton rate to follow changes
is approximately equal to that of the spectral function. The
quantum mechanical nature of the regarded systems leads to
oscillations and negative values in the changing spectral
functions, occupation numbers and production rates as well as
interferences that one does not get in semi-classical, adiabatic
calculations. The "rate" calculated here possesses the full
quantum mechanical information incorporated and contains "memory"
interferences that might cause cancellations - hence the rate has
to be able to become negative while the time integrated yield
always stays positive as the only observable physical quantity. An
intriguing example for the occurring oscillations is shown in Fig.
\ref{fig:oscillations}.
To calculate the yield, we model the fireball evolution and fold it
with the calculated time dependent rates, similar to \cite{Rapp:1999ej}.
For the effective volume we choose a longitudinal Bjorken expansion together with an
accelerating radial flow
\begin{equation}
    V_{eff}(t\geq t_0)=\pi c t (r_0+v_0(t-t_0)+0.5 a_0 (t-t_0)^2)^2\text{,}
    \label{veff}
\end{equation}
with $r_0=6.5$ fm, $v_0=0.15~c$ and $a_0=0.05~c^2/\text{fm}$ (see
also \cite{Greiner:2001uh}). From (\ref{veff}) and the constraint
of conserved entropy (given by a constant entropy per baryon
$S/A=30$ for SPS energies \cite{Greiner:1991us}), temperature
$T(\tau)$ and chemical potentials follow as functions of time. We
start the calculation at the freezeout temperature of 175 MeV,
whereas the final temperature, reached after a lifetime of about
7.8 fm/c is 120 MeV (thermal freezeout). At this point, we turn
off further dilepton production by a rapid decrease of the
temperature towards zero. With the time dependent temperatures we
can integrate the rate and immediately find the yield per unit
four momentum. The results for different scenarios are shown in
Fig. \ref{fig:yieldmassres}. We compare to Markov calculations
that assume instantaneous adaption of the spectral function (and
the rate) to the self energy, as employed in \cite{Rapp:1999ej}
(dashed lines in Fig.\ref{fig:yieldmassres}). The yield resulting
from a mass shift following from assumed Brown-Rho scaling
\cite{br91} shows an enhancement of about a factor of 2 in the
dynamic calculation within the mass range where the CERES
experiment \cite{ce95,Wessels:2002ha} has measured a strongly
enhanced dilepton yield compared to calculations with vacuum
spectral functions. Also the coupling to the N(1520) resonance
with no broadening shows an enhanced production around the
resonance peak and the $\rho$ vacuum peak, but also a reduced
yield in the region between the peaks. On the other hand, the
difference to Markov calculations becomes smaller for large
in-medium widths.


\section{Summary and Conclusions}
\label{conclusion}
    In the present work we introduced a method to calculate dilepton
    production rates within a non-equilibrium field theory formalism,
    based on the real time approach of Schwinger and Keldysh.
    We investigated possible medium modifications of the $\rho$
    meson in a fireball created in a heavy ion
    collision. We considered mass shifts, broadening and coupling
    to resonances.
    Special attention was put to possible retardation effects
    concerning the off-shell evolution.

    The timescale on that the spectral function adjusts to changes in the self energy was found to be proportional to the inverse vacuum width of the
    meson $\Gamma_2$ like $c/\Gamma_2$, with $c$ approximately 3. Further dependence on the in-medium width
    as well as on the size of the medium modification is present.
    We find typical retardations for the $\rho$ of $3$ fm/c and about $60$ fm/c for the $\omega$, a \emph{very slow} adjustment.

    The full quantum field theoretical treatment leads to
    oscillations in all mentioned quantities when changes in the self energy are performed.
    This oscillatory behavior reveals the quantum mechanical
    character of the many particle system, present in the investigated heavy ion reaction.
    The oscillations potentially cancel when the rate is integrated over time such that the measurable
    dilepton yield is always positive.

    Comparison of dynamically calculated yields with those calculated assuming adiabaticity reveals differences.
    About a factor of $2$ difference was found within the invariant mass range of $250$ to $500$ MeV
    for mass shifts predicted using Brown-Rho scaling for the $\rho$-meson in a fireball at SpS energies (158 AGeV).
    This is the range where CERES measures an increased dilepton
    yield as compared to calculations assuming the $\rho$'s vacuum
    shape. Similar results were found for the coupling of the $\rho$-meson to resonance-hole pairs.
    Our findings show that exact treatment of medium modifications in principle requires the consideration of memory
    effects.
    Further investigation of the $\omega$-meson is being done - due to its small width it causes numerical complications.
    It has to be seen how our results can be consistently
    (and probably in an approximative manner) incorporated in semi-classical nonequilibrium transport codes
    like UrQMD \cite{Bass:1998ca}, BUU \cite{Effenberger:1999ay} or HSD \cite{Cassing:1999es}.\\

\bibliography{schenke}

\end{document}